\documentclass[12pt]{article}
\usepackage{geometry}                
\usepackage{graphicx}
\usepackage{amssymb}
\usepackage{epstopdf}
\def\0#1{{\mathrm{#1}}}
\def\1#1{{\mathbb{#1}}}
\def\2#1{{\mathbf{#1}}}  
\def\3#1{{\mathcal{#1}}}
\def\4#1{{\mathsf{#1}}}   
\def\5#1{{{\widetilde{#1}}}}  
\def\6#1{{\overline{#1}}} 
\def\7#1{\breve{#1}}
 \def\8#1{{\widehat{#1}}}  
 \def\9#1{{\widecheck{#1}}}
 
 \def\cir#1{{\stackrel{\circ}{#1}}}

 \def\<{{\left<\right.}}
\def\>{{\left.\right>}}
 
 \def\ox{\otimes}

\def\x{\times}

\def\BAR{\begin{array}}
\def\EAR{\end{array}}
\def\BEQ{\begin{equation}}
\def\EEQ{\end{equation}}
\def\BEN{\begin{enumerate}}
\def\EEN{\end{enumerate}}
\def\BIT{\begin{itemize}}
\def\EIT{\end{itemize}}
\def\BEA{\begin{eqnarray}}
\def\EEA{\end{eqnarray}}
\def\BED{\begin{description}}
\def\END{\end{description}}
\def\BET{\begin{table}}	
\def\ENT{\end{table}}

\def\apo{\mbox{\bf '}}
\def\cliff{\mathop{{{\sqcap}}}\nolimits}
\def\Cliff{\mathop{{\mathsf{Cliff}}}\nolimits}
\def\grade{\mathop{{\mathsf{grade}}}\nolimits}

\def\dual{\mathop{{\mathsf {dual}}}\nolimits}

\def\dup{\mathop{{\mathsf {dup}}}\nolimits} 

\def\hexp{\mathop{{\mathrm{hexp}}}\nolimits}

\def\op{{\mathop{\mbox{op\,}}\nolimits}}

\def\oar{\rightharpoonup}

\def\Qi{{\cir{\imath}\,}}
\def\so{\mathop{{\mathsf {so}}}\nolimits}
\def\SO{\mathop{{\mathsf {SO}}}\nolimits} 

\def\spin{\mathop{{\mathsf {spin}}}\nolimits}

\def\W{{\bigwedge}}
\def\w{\mathop{\wedge}\nolimits}

\def\Z{{ \vrule width2pt height0pt depth0pt}}

\newtheorem{proposition}{Proposition}
\newtheorem{hypothesis}{Hypothesis}

\title{Quantum set algebra for quantum field theory}

\author{David Ritz Finkelstein\\
\small Georgia Institute of Technology, Atlanta, Georgia\footnote{Emeritus.}\\
\small finkelstein@gatech.edu}

 \date{}  

\begin{document}
\maketitle

\begin{abstract}
 Quantum field theory can be physically regularized
by modularizing it on several levels  of aggregation.
Since computation is already thoroughly modularized,
 physical experiments are treated here
as quantum relativistic cellular computations
with spins for cells, address, memory, and control registers.
For regularity the modules are taken to be iterated Fermi-Dirac assemblies.
These are shown to be spins 
in various dimensions.
Bose statistics are expressed as approximations
to the Palev statistics of pairs of Fermi-Dirac quanta.
\end{abstract}


\maketitle


\section{{Why modularize} \label{S:Why}}

The main goal of this study is a physical regularization of relativistic quantum field theory, including the Standard Model and gravity, as opposed to a formal regularization.
The terms ``physical" and ``formal"  come from \cite{PAULI1949}.

The von Neumann axiomatization of quantum theory 
in Hilbert space was abstracted from Heisenberg's non-relativistic theory of the hydrogen atom and is itself deeply non-relativistic.
Relativistic locality makes each point of a spacelike surface a
separate physical system, 
in that the variables it carries are independent of those of any other point
of the surface.
Therefore a relativistic theory is necessarily a many-system theory.
In a quantum theory this implies an algebra of creation and annihilation operators, not a mere vector space.
The Hilbert space theory is a one-system theory.
Its constructs correspond to those of classical predicate
algebra, with no analysis into independent systems.

In compensation, the practice has been to multiply an infinite number of infinite-dimensional Hilbert spaces into a Hilbert space for a quantum field system, a double infinity for which there is no physical evidence.
Here the standard Hilbert space
quantum theory is not multiplied but divided
into a finite number of finite-dimensional relativistically invariant local quantum systems.

In the early 20th century there was great resistance to replacing the state-based
physics with an operation-based physics.
Some state-based ideas persist into the quantum era today.
To avoid them
it is convenient to regard a physics experiment with an input and output as
a computation carried out by the experimenter 
on the experimentee,
which
accepts the input and delivers the output.
The experimenter is the user and the experimentee is the computer
\cite{FINKELSTEIN1969a}.
The experimenter has no access to any state of the experimentee,
only to her or his input and output operations,
just as the user of a laptop has no access to its state.

In such a computer formulation of classical field theory,
space-time is the address system,
the field is the  distributed memory, 
and both are analogue systems.
Again, a small fast quantum computation occurs in every particle collision event.
Here physical experimentation is modeled as finite quantum relativistic computation.

A computer that merely solves the equations of a physical theory
is not relevant to this study.
The inputs and outputs to the physical computation studied here
are not data about quanta, but quanta themselves.
The computation time is physical time
 and every experiment has a definite outcome,
 even though quantum theory only predicts the outcome 
 of a set of experiments of measure 0.
The fundamental elements are not data
but acts of input and output of physical quanta, 
also called creation and annihilation.
The syntax is a quantum variant of finite set theory,
with physical meanings.
This assures finiteness from the start.

There is also a gain in self-consistency. 
Set theory, now the lingua franca of mathematics, 
and therefore of mathematical physics,
rests on the predicate algebra of Boole. 
Quantum theory, however, uses a revised
predicate algebra for its assertions about a quantum system, 
a quantum logic \cite{NEUMANN1932}.
This means, for example, 
 that we cannot meaningfully speak of a set
 of electrons in the sense of Cantor.
Here 
a quantum set is defined as a Fermi-Dirac or {\em odd} assembly.
Then a quantum set of electrons is a well-defined familiar
construct.

The classical algebra of finite sets
is {\em multiordinal}\/---refers to several levels of aggregation at once
(a coinage of Korzybski)---and {\em autogenic}\/: 
generated by its own operations  of 
multiplication $\w$
and association $\{\dots\}$.
Multiplication is the disjoint union;
association is
called unitization by Dewey.
This is enough to formulate finite classical
computations without variables.
Therefore a multiordinal autogenic Grassmann algebra $\3S$ 
of finite quantum sets is defined and used to formulate similar
quantum computations. 
Here elements of $\3S$
represent histories of quantum experiments 
or quantum computations.

Although $\3S$ is infinite-dimensional, 
no cut-off is needed;
quantum physics simply does not use quantum sets in $\3S$ of rank higher than 7 so far.
In contrast, the classical continuum is a classical set of infinite rank.

The computation models considered are quantum, relativistic,
and autogenic
\cite{FINKELSTEIN1996}.
Other quantum computational models of Nature 
have been studied by
G. M. D'Ariano \cite{DARIANO2011} and S. Lloyd \cite{LLOYD2006}.
Non-quantum non-relativistic models  
have also been advanced, for example by K. Zuse, E. Fredkin, and S. Wolfram.

\section{Ports\label{S:PORTS}}

In the Standard Model or any other quantum field theory, 
and therefore in these computer models,
an experiment is a network of 
operations of quantum creation and annihilation, 
or input and output.
For brevity, call these 
{\em port operations}, portations, or most briefly ports, 
and say that they {\em import} or {\em export} quanta.

Boole's {\em Laws of Thought} 
 and the set theory of Cantor, in which
``A set is a Many that allows itself to be 
thought of as a One," explicitly concern mental processes.
Adapted to quantum physical processes, 
the Laws of Thought of Boole 
become Laws of Ports and the set of Cantor becomes
a Many that can be ported as One.

The algebra $\3S$ is autogenic in that its space is 
not posited entire as Hilbert space is,
but,  like finite set theory $S$, 
is built up from nothing, tier by tier, 
by its own algebraic
operations $+, \x, \{\dots\}, \1R$.

A subspace of $\3S$ cannot support a finite-dimensional
faithful unitary representation of the Lorentz group,
since the finite-dimensional unitary groups are compact
and the Lorentz group is not.
I sacrifice unitarity
for the Lorentz invariance and the finite-dimensionalality of spins,
and recoup it, hopefully, in the passage to
an infinite number of spins.

Port operators and vectors describe 
the experimenter's modi operandi, 
not the system's status quo \cite{WIGNER1963}.
Every experiment has a beginning and an end,
and therefore has both import and export operations,
mutually dual,
each defined relative to the other,
in the past and future of the interactions under study.

The common name ``state" 
makes this duality seem to be a problem: 
How can a system have two states?
The term ``port" avoids this false problem: 
How can an experiment not have an input and an output?

To address this terminological problem 
Giles redefined ``state" to mean import operation
and calling export operations {\em tests} \cite{GILES1970}.
But this redefinition has not caught on.
It we continue to understand 
a state as a complete description,
then according to quantum theory
a quantum system has no state. 

\section{Quantum set algebra\label{S:S}}

A quantum set is an odd assembly of odd assemblies of $\dots$, 
founded on the empty set.
The quantum set algebra $\3S$ 
is the semigroup of finite-rank sets,
generated by $1, \w,$ and $\iota=\{\dots\}$,
with familiar syntax and postulates.
$\3S$ is the minimal real Grassmann algebra 
that is closed under a suitably linearized variant 
of the Cantor-Peano associator $\iota: x\mapsto \{x\}$.
$\iota$ becomes a free linear operator mapping $\3S$ onto 
its first grade $\3S_1$.
Thus $\3S$ is its own Grassmann algebra; more exactly,
$\3S$ is the Grassmann algebra over its own associations: 
\BEQ \label{E:SWS}
\3S=\W\iota\apo \3S\/.
\EEQ

About operational meanings: 
Grassmann addition $+$ of $\3S$ represents quantum superposition.
Grassmann
multiplication $\w$ of $\3S$ represents concatenation
subject to the Pauli Exclusion Principle $\iota x\w \iota x=0$.
I do not know the operational meaning of association $\iota$ yet,
but it is used implicitly in physics since Newton,
for example in the concept of function of time,
which associates values with instants of time.
I use it only in the context of (\ref{E:SWS}).
Von Neumann took the concept of function as fundamental in his set theory. 
$\iota$ can be regarded as a great restriction of that concept, to functions with range $2=\{0,1\}$.
In the absence of an operational meaning,
the decision to use Cantor's $\iota$ is somewhat arbitrary.

Since Chevalley uses a Grassmann algebra for simplicial complexes,
we may call a quantum entity with port vectors in $\3S$
a {\em quantum complex} as well as a quantum set, 
with the understanding that each vertex $x$
of this complex is the association $x=\iota y$ of some simplex 
$y$, or a superposition of such associations.
The system (with ports in) $\W \3V$ is a quantum complex 
of a quantum simplex (with ports in) $\3V$.

Quantum field theory
has as its addresses 
classical space-time variables 
with infinite spectra,
and its ports carry infinite information 
(negative infinite von Neumann entropy).
In contrast,  the $\3S$ process has quantum sets for its quantum addresses and memories,
its variables have finite spectra,  
and its ports carry zero entropy,
which can at least be approximated cryogenically.

\paragraph*{\bf Notation \vrule width 0pt depth 0pt height 12pt}

$x\w y$ represents the disjoint union of $x$ and $y$.

$\{x,y,\dots\}= \iota x \w \iota y \w \dots$.

In Pierce's theory of Boolean algebra,
$x\dotplus x = \infty$.
In the notation $\3S$ 
this becomes  the Pauli Exclusion Principle $x\w x=0$,
equivalent to $x\w y=-y\w x$.

$\3S$ is the Grassmann algebra of port vectors for the q set,
and is generated by $\1R, \w,\iota$\/, and $+$,
with the usual syntax and postulates.
$\3S$ generates its own coefficients; I provide $\1R$ to save time.

Let $\3X\subset \3S$ be a subspace of $\3S$. 
The vector space $\iota\apo \3X$ is
\BEQ  
\iota\apo \3X:=\{y=\iota x|\; x\in \3X\}
\EEQ
(``the associations of $\3X$").

1 is the empty set and has rank 1. $1\w x= x$.

The port vector $0$ is a space-holder meaning nothing,
like the vector 0 in a ket space and
unlike the empty-set port vector 1,
which means ``Nothing", the Fock-space vacuum.

$\grade:\3S\to \3S$ is a linear operator such that 
for homogeneous polynomials $\psi\in \3S$ of degree
$g$ in unit sets, $\grade \psi=g\psi$.
Then 
\BEQ
\grade AB = (\grade A)B + A(\grade B).
\EEQ

$g$-adics (monadics, dyadics, $\dots$) are Grassmann or Clifford elements of grade $g$.

${\grade}_g\3A$ stands for the grade-$g$ subspace 
of the graded algebra $\3A$. 

$\3S(r)$ is {\em rank} $r$ of $\3S$,
spanned by polyadics in $\3S$
with no more than $r$ nested applications of $\iota$.

$\Delta\3S(r):=\3S(r)- \3S(r-1)$ is {\em tier} $r$ of $\3S$, 
the complement in $\3S(r)$ of $\3S(r-1)$.

$\3S(R|r):=$ {\em rank $R$ modulo rank $r$:=} the subspace of $\3S(R)$
spanned by basis elements of $\3S(R)$ with no
factor of rank $r$ except 1. 
The union of the $R-r$ tiers 
$$\Delta\3S(R), \Delta\3S(R-1), \dots, \Delta\3S(r+1).$$

$\iota x$, $\{x\}$, and sometimes $\6x$ all designate
the monad formed from $x$ by association.
For example 
\BEQ
\6{{\6{\61}\,{\61}}}=\{\{\{1\}\}\{1\}\}.
\EEQ
$\iota$ is defined so that $ \forall A, B\in \3S, \lambda \in \1R$,
\BEA
\iota(A+\lambda B)&\equiv &\{A\} +\lambda \{B\},\cr
\grade \iota x &=& \iota x.
\EEA

The {\em hyperexponential} function $\hexp r$ is the integer defined for $r\in \1N$ by
\BEA\label{E:HEXP}
\hexp 0&:=&1, \cr
\hexp (r+1) &:=& 2^{\hexp r} \quad \mbox{for } r\ge 0.
\EEA

Sub- indicates an inclusion relation; 
pre- a possibly iterated membership relation.
Thus $x$ is a pre-spin of the spin $\iota\iota x$.

Each entry in Table \ref{T:QSETS} represents both a state of the finite-rank classical set
and a basis vector of the port vector space $\3S$ of the quantum set,
of rank $r$.
The table is complete for rank 4
and is but a small sample of rank 6, whose
full text does not fit into the visible universe.
\begin{table}[h!]
\begin{center}
\begin{tabular}{|r|ccccccccccccc|}
\hline
$\vdots$ &&&& && &$\vdots$&&&& &&\cr
\hline
6 & \vrule height 18pt depth 0pt width 0pt
$\6{\6{\6{\6{\6{\6{\Z}}}}}}$&
$\6{\6{\6{\6{\6{\6{\Z}}}}}}\,{\6{\Z}}$&
$\6{\6{\6{\6{\6{\6{\Z}}}}}}\,{\6{\6{\Z}}}$&
$\6{\6{\6{\6{\6{\6{\Z}}}}}}\,\6{\6{\Z}}\,\6{\Z} $&
$\6{\6{\6{\6{\6{\6{\Z}}}}}}\,\6{\6{\6{\Z}}}$&
$\6{\6{\6{\6{\6{\6{\Z}}}}}}\,\6{\6{\6{\Z}}}\,\6{\Z}$&
$\6{\6{\6{\6{\6{\6{\Z}}}}}}\,\6{\6{\6{\Z}}}\,\6{\6{\Z}}$&
$\6{\6{\6{\6{\6{\6{\Z}}}}}}\,\6{\6{\6{\Z}}}\,\6{\6{\Z}}\,\6{\Z}$&
$\6{\6{\6{\6{\6{\6{\Z}}}}}}\,\6{\6{\6{\Z}}\,\6{\Z}}$&
${}{\cdots}$&
$\6{\6{\6{\6{\6{\6{\Z}}}}}\,\6{\Z}}$&
$\6{\6{\6{\6{\6{\6{\Z}}}}}\,\6{{\Z}}}\,\6{\Z}$&
${}{\cdots}$\cr
\hline
5 & \vrule height 14pt depth 0pt width 0pt
$\6{\6{\6{\6{\6{\Z}}}}}$&
$\6{\6{\6{\6{\6{\Z}}}}}\,\6{\Z}$&
$\6{\6{\6{\6{\6{\Z}}}}}\,\6{\6{\Z}}$&
$\6{\6{\6{\6{\6{\Z}}}}}\,\6{\6{\Z}}\,\6{\Z} $&
$\6{\6{\6{\6{\6{\Z}}}}}\,\6{\6{\6{\Z}}}$&
$\6{\6{\6{\6{\6{\Z}}}}}\,\6{\6{\6{\Z}}}\,\6{\Z}$&
$\6{\6{\6{\6{\6{\Z}}}}}\,\6{\6{\6{\Z}}}\,\6{\6{\Z}}$&
$\6{\6{\6{\6{\6{\Z}}}}}\,\6{\6{\6{\Z}}}\,\6{\6{\Z}}\,\6{\Z}$&
$\6{\6{\6{\6{\6{\Z}}}}}\,\6{\6{\6{\Z}}\,\6{\Z}}$&
${}{\cdots}$&
$\6{\6{\6{\6{\6{\Z}}}}\,\6{\Z}}$&
$\6{\6{\6{\6{\6{\Z}}}}\,\6{{\Z}}}\,\6{\Z}$&
${}{\cdots}$\cr
\hline
4 & \vrule height 14pt depth 0pt width 0pt
$
\6{\6{\6{\6{\Z}}}}$&$
\6{\6{\6{\6{\Z}}}}\,\6{\Z}$&$
\6{\6{\6{\6{\Z}}}}\,\6{\6{\Z}}$&$
\6{\6{\6{\6{\Z}}}}\,\6{\6{\Z}}\,\6{\Z} $&$
\6{\6{\6{\6{\Z}}}}\,\6{\6{\6{\Z}}}$&$
{{\6{\6{\6{\6{\Z}}}}\,{\6{\6{\6{\Z}}}\,\6{\Z}}\,}}$&$
{{\6{\6{\6{\6{\Z}}}}\,\6{\6{\6{\Z}}}\,\6{\6{\Z}} }}$&$
{{\6{\6{\6{\6{\Z}}}}\,\6{\6{\6{\Z}}}}\,\6{\6{\Z}}\,\6{\Z}}$&$
\6{\6{\6{\6{\Z}}}}\,\6{\6{\6{\Z}}\,\6{\Z}}\,\6{\Z}$&$
{}{\cdots}$&$
{{\6{\6{\6{\6{\Z}}}\,\6{\Z}}} }$&
${{\6{\6{\6{\6{\Z}}}\,\6{\Z}}}}$&$
{}{\cdots}$
\cr
& \vrule height 14pt depth 0pt width 0pt
16&17&18&19&20&21&22&23&24&\dots&32&33&\dots
\cr
\hline
3 & \vrule height 14pt depth 0pt width 0pt
$
{\6{\6{\6{\Z}}}}$&$
{{\6{\6{\6{\Z}}}\,\6{\Z}}}$&$
{{\6{\6{\6{\Z}}}\,\6{\6{\Z}}}}$&$
{{\6{\6{\6{\Z}}}\,\6{\6{\Z}}\,\6{\Z}}}$&$
{{\6{\6{\6{\Z}}\,\6{\Z}}}}$&$
{{\6{\6{\6{\Z}}\,\6{\Z}}\, \6{\Z} }}$&$
{\6{\6{\6{\Z}}\,\6{\Z}} \, \6{\6{\Z}}\,}$&$
{\6{\6{\6{\Z}}\,\6{\Z}}\,\6{\6{\Z}}\,\6{\Z}}$&$
{\6{\6{\6{\Z}}\,\6{\Z}}\,\6{\6{\6{\Z}}}}$&$
{\,\6{\6{\6{\Z}}\,\6{\Z}}\,\6{\6{\6{\Z}}}\,\6{\Z}}$&$
{\6{\6{\6{\Z}}\,\6{\Z}}\,\6{\6{\6{\Z}}}\,\6{\6{\Z}}}$&$
{{\6{\6{\6{\Z}}\,\6{\Z}}\,\6{\6{\6{\Z}}}\,\6{\6{\Z}}\,\6{\Z}}}$&
\cr
& \vrule height 14pt depth 0pt width 0pt
4&5&6&7&8&9&10&11&12&13&14&15&
\cr
\hline
2 & \vrule height 14pt depth 0pt width 0pt ${\6{\6{\Z\Z}}}$&$
{{\6{\6{\Z\Z}}\,\6{\Z\Z}}}$& &&&&&&&&&&\cr
& \vrule height 14pt depth 0pt width 0pt 2&3& &&&&&&&&&&\cr
\hline
1 & \vrule height 14pt depth 0pt width 0pt${\6{1}}$&
&&&&&&&&&&&\cr
 & \vrule height 14pt depth 0pt width 0pt${1}$&
&&&&&&&&&&&\cr
\hline
0 & \vrule height 14pt depth 0pt width 0pt  1&
&&&&&&&&&&&\cr
 & \vrule height 14pt depth 0pt width 0pt  0&
&&&&&&&&&&&\cr
\hline
\hline
\vrule height 12pt depth 0pt width 0pt $r$
&&&&&&$$&$e^n$&&&&&&
\cr
\vrule height 12pt depth 0pt width 0pt 
&&&&&&$$&$n$&&&&&&
\cr
\hline
\end{tabular}
\end{center}
\caption { Basis vectors of the quantum set}
\label{T:QSETS}
\end{table}
The ``1" that is explicit in $e^0$ and $e^1$ 
is implicit beneath each bottom bar of the other basis vectors $e_n$,
and represents the empty set.
Serial numbers are given under the first 25 elements.
In one model tiers 0--2 correspond to the four leptons and tier 3 to the 12 quarks.

The grade of a set is the number of top bars in its width.
The rank of a set is its height in bars.
Fan-in occurs in every figure with fewer top bars than bottom bars.
Fan-out is the recurrence of one figure inside
others of higher rank.
$S$ and $\3S$ are fractal in that 
every element supports an isomorph of the whole net.

$\3S(4)$ has $\hexp 4=65,536$ elements, not enough for space-time,
but $\3S(5)$ has $\hexp 5=2^{65,536}$, probably
more than enough for all possible physical events.
Thus despite our present ignorance, we can be fairly sure that
the transition from microscopic to macroscopic
happens between rank 4 and rank 5.
Then $\3S(6)$ suffices for all possible physical histories,
sets of events.

In the Standard Model, 
the fundamental fermions have no parts, 
but they have elements, such as spin and isospin.
The Grassmann skew-symmetry postulate 
extends the Pauli Exclusion Principle 
from the fundamental fermions to
to their modules of any lower rank.

A $\1C$ quantum theory is a specialization 
of an $\1R$ quantum theory
that singles out one real operator $\Qi$ 
with $(\Qi)^2=-1$ 
and requires observables to commute with $\Qi$
\cite{STUECKELBERG1960}.
Complex port rays correspond to real $\Qi$-invariant planes.

The  operator algebra $\3T=\op \3S$ is  generated by
odd import and export operators and is the Clifford algebra
of a neutral metric.
$\3T$
contains the observables or q numbers and the
dynamical development operators of the $\3S$ process.
I assume that $\3T$ includes the Clifford algebras
of the internal variables of the Standard Model,
much as proposed by Wilczek and Zee \cite{WILCZEK1982},
but also the orbital variables.

The following elementary relation is useful for counting 
the replicas of a low-rank cell in a high-rank complex.
The same relation holds among classical sets, where the factor order does not matter and there are no signs.

{\proposition [Replication] 
$\forall r_1 > r_2>\dots>r_n  \in \1N:$
\BEQ
\3S(r_1) =\3S(r_1|r_2)\w \3S(r_2|r_3)\w \dots \w\3S(r_n)
\EEQ
}
\noindent{\bf Proof}. Analogous to the classical case.
$\square$

\section{Binary and hyperbinary numbers\label{S:HYPER}}

Any finite-rank set has an integer {\em serial number} telling the order of its generation, set at 0 for the port vector 1
of the empty set.
The {\em hyperbinary expansion} of a natural number $n$
is the
sequence 
\BEQ
n=\4h[h_r h_{r-1}\dots h_1] := \sum_{r=0}^{\infty} h_r \hexp(r).
\EEQ
The value of 1 in the $r$th place is
the exponential $2^{r}$ in the binary expansion,
and is
the hyperexponential $\hexp r$ in the hyperbinary.
 
The number beneath some of the symbols in Table \ref{T:QSETS} 
is the {\em serial number},
giving the order of generation,
counting  from left to right from bottom to top,
and
starting from the empty set $1$ with serial number 0.
$e^n$, the symbol with serial number $n$,
can be regarded as a hyperbinary symbol for $n$.
Tier $r$ consists of the numbers with $r$ hyperbinary places.
If $e^n$,  $e^{n'}$, and $e^{n''}$ are basis elements in $\3S$
and $e^{n'}e^n=e^{n''}$ 
 then $n''=n'+n$.

A stack of $r$ bars is $\hexp r$, the value of a 1 in hyperbinary place $r$.
If $H\in S$ is a set then $\6H=2^H$. 

In the generalization to hyperbinary q numbers,
each place holds an operator with the given spectrum of values. 

Each hyperbinary number $n$ thus labels a set in $S$, 
a basis ket 
$\<n| \in \3S$, and a 
bra $|n\>\in \dual \3S$: the $n$th entry in Table I.
The $\<n|$ define the spectral resolution of the q address,
\BEQ
\2n=\sum_n |n\> n\<n|\/.
\EEQ

For example, the quantum simplicial computations of particle physics require seven-figure hyperbinary numbers, using places 0---6.

\section{Quantization\label{S:Q}}

Segal 
noted that both relativization and quantization 
made the Killing form
of physical Lie algebras less singular \cite{SEGAL1951}.
He inferred
{\hypothesis [Segal] 
Physics evolves toward regular Lie algebras.}

Perhaps this is because these are exactly the ones
that are stable up to isomorphism 
under small errors in the structure tensor.
Take this hypothesis as a provisional road sign.

Then both
relativization and canonical quantization
are unfinished.
They make
singular groups of the prior theories
less singular, 
but not yet regular.
General relativization introduces regularizing non-commutativities like curvature
but also gauge singularities.

Call a regularizing deformation a {\em re\-formation}.
Yang's re\-formation of the 15-dimensional Heisenberg-Poincar\'e algebra
$\4a(x,p,L,i)$ is a 15-dimensional $\so(5,1)$ algebra
with two new space-time dimensions and two new fundamental constants, 
which can be chosen as a quantum time unit $\4T$ and 
a large integer $\4N$, the magnitude of an angular momentum 
\cite{YANG1947}.
The Yang theory also implies a new self-organization that freezes out the two 
new space-time dimensions at low energies.
This too
was still not a true regularization,
since he used an infinite-dimensional unitary representation.
The $\3S$ model given below completes this regularization and others.
It simply represents Yang's Lie algebra with spin operators, which are regular,
instead of differential operators, which are singular.

The
stable structures of the Standard Model need no reform.
Therefore $\3S$ models inherit the odd
 statistics and the
spin, 
isospin, and color groups of the Standard Model.
But they must reform the Poincar\'e group, Bose statistics,
hypercharge,  bundle gauge groups, 
and the unitary group of infinite-dimensional Hilbert space. 

An $\3S$ model reforms the usual continuum differential geometry in 
two steps.

1.  Translations become rotations in planes normal to Minkowski space-time. 
First-grade coordinate infinitesimals $dx^{\mu}$ 
and the imaginary $i$
become second-grade spin components $\gamma^{y'y}$.
Integrals become sums.
  
2. It expresses all spin operators $\gamma^n$ 
as port operators for lower-rank elements,
{\em pre-spins}.

That Dirac spins underlie Minkowski space-time I heard from Feynman \cite{FEYNMAN1941}.
It is a relativistic form of an idea I had earlier heard from 
Roger Penrose \cite{PENROSE1971}.
The grade-doubling of step 1 is that of Yang \cite{YANG1947} and Segal \cite{SEGAL1951}.
It replaces
the canonical Lie algebra by a spin Lie algebra,
and canonically conjugate pairs by complimentary spin triples $\2S=\{S_1, S_2, S_3\}$
in the relation $\2S\x\2S\sim\2S$.

That spin operators are port operators gives quantum-physical meaning to Cartan's original theory of spinor spaces as Grassmann algebras:
A spin is an odd assembly of lower-rank elements, 
pre-spins, whose  port vectors are Cartan's semivectors.

As a result the space-time and energy-momentum coordinates of $\3S$ models
all have finite discrete spectra
but are still  invariant under a group that 
is nearly Poincar\'e.

The quantum logic of {$\3S$}  is still a projective geometry.
It is endowed, however, with a polarity 
defined by odd statistics, 
a neutral (or ``split" \cite{KOSTANT1986})
 quadratic
form, the duplex form of Section \ref{S:MOD}.
This microscopic neutral metric constricts to a macroscopic
Minkowskian one by a hypothetical self-alignment of spins
in one plane
that freezes out two extra variables
as it forms the imaginary unit $i$ \cite{YANG1947}.
I define $\3S$ to be a real Grassmann algebra
because the usual $i$ is generated in this way.

Since
the interaction vertex contributes to propagation,
a parsimonious dynamics is conceivable in which
both are composed from the same vertex.

Odd statistics is stable but Bose is not.
The original canonical quantization left out odd variables altogether.
Supersymmetric quantum theory remedies this by
giving odd and even variables equal footing,
with a ``particle-ino" for every Bose particle
and hopes for a cancellation of infinities.
Quantum set theory discards basic even statistics 
entirely and reconstructs it from odd.
Pairs of odd $\3S$ quanta have exact (and stable) Palev statistics based on a spin algebra (Section \ref{S:PALEV}).
Then Palev statistics has Bose statistics as a singular  limit with self-alignment.

This relativistic derivation of Bose statistics from Fermi
resembles non-relativistic ones of Bohm, Tomonaga \cite{TOMONAGA1950}, Luttinger, Mattis and Lieb \cite{MATTIS1965}, 
and others.
\cite{MATTIS1965} is a key paper for the non-relativistic process
and provides
valuable references. 

Therefore $\3S$ models avoid the Bose instability by
dropping fundamental bosons at all levels
and synthesizing even particles out of odd.
As a result, in an $\3S$ model of light no
two photons are exactly identical.

\section{Spins as Modules\label{S:MOD}}  

Modular architectures have been needed to cope with complexity 
in industry, computers \cite{SIMON1962}, finance, government, and
 biological systems \cite{KOZO2010, MARGULIS2002}. 
This suggests
{\hypothesis [Modularity ] \hfil\\
\rule{6pt}{0pt} Nature has an autogenic modular architecture.}
\vskip8pt
That is, Nature generates itself in stages by processes of Nature, each stage an assembly of
modules generated in earlier stages.
The operation that is crucial for this is association $\iota$.

The Modular Hypothesis includes the Atomic Hypothesis and extends it from matter
 to space-time.
In quantum theories the modules are represented by port operations
that import or export them.

Import and export operations are now associated with rays in
the same vector space $\3S$
so that a transition $A\to B$ is forbidden, 
$B\leftarrow\kern-10pt | \kern7pt A$,
if the scalar product of the vectors of one ray with those of the other are all 0: 
\BEQ B\leftarrow\kern-9pt | \kern7pt A \quad :\equiv:\quad \forall a\in A, b\in B:\quad b\cdot a = 0.
\EEQ  
Since transitions from one importation to another do not occur,
the import vectors of any frame are null (isotropic) vectors 
of the scalar product, and so are the export vectors.

Yang indicates how an imaginary unit $i$ emerges from
a simple real quantum space-time algebra
as a component of higher-dimensional angular momentum (\cite{YANG1947}).
Therefore I assume $\3S$ has the coefficient field $\1R$.
This implies that 
at least under extreme conditions to be specified, 
it is physically possible
to distinguish between
ports $\psi$ and $i\psi$,
usually considered indistinguishable.

For an autogenic modular architecture, 
I endow $\3S$ with
a linearized operation of association
$\iota \psi=\{\psi\}$.
This is already used widely but implicitly, as follows.

The Standard Model (augmented with dextral neutrino)
uses modules of
orbit, spin, hypercharge, isospin, color, and generation,
with defining port vector spaces 
$\3O, \3W, \3Y, \3I, \3C,\3G$, respectively;
where in one option $\3W\cong 2\1C$ is Weyl spinor space.
These modules are elements, not parts, of the fermion:
\BEQ
\3F= \{\3O\w \3W\w \W\3I \w \W\3C\ox \dots\},\quad
\3O\not\subset \3F, \quad \3O\in\3F\/.
\EEQ
Here $\W \3I$ is the Grassmann algebra over $\3I$.
It allows for the fermions that have no isospin.
Similarly for $\W\3C$ allows for the fermions without color.

If a two-fermion port space 
\begin{eqnarray}
 \3F\x \3F'= \{\3O\w \3W\w \W\3I \w \dots\}\w\{\3O\w \3W\w \W\3I \w \dots \}
\end{eqnarray}
were written
without the associating braces,
the two fermions would be scrambled.

In the general case
\BEQ
\{a\w b\}\w c\ne a\w\{b\w c\}.
\EEQ
Association interrupts associativity.

$\3S$ models are  modular from the start.
The modules are unit q sets with port vectors in the modular Grassmann algebra $\3S$.
The association of a product of modules of one rank
is a module of the next rank 
that transforms like their product 
but may enter as a unit into further modules.

A {\em spin} is a quantum system whose operator algebra 
is a Clifford algebra. 
Its port vectors or spinors are then the column vectors 
(or minimal left ideals) of that algebra.
Spins of the Lorentz group and isospins are special cases.
According to $\so(10)$ GUT, quark color is another.

{\proposition All quantum sets are spins.}

\strut\\
{\bf Proof:}  For any vector space $\3V$, define
the {\em duplex space} $\3W:=\dup \3V:=\3V + \dual \3V$  
with the {\em duplex quadratic form} $\4D:\3W\to \dual \3W$ given by
\BEA
\forall v\in\3V,\, v'\in \dual \3V,\, w&=&v+v'\in \3W:\cr\quad \|w\|_{\4D}:=w \4D w :=\|v+v'\|&:=& v'\circ v\/.
\EEA
Let the module port space be
\BEQ
\3V'=\W \3V\/.
\EEQ
 Then
\begin{eqnarray}
\op \3V' &=&\Cliff [\3V+\dual \3V]\cr
&=&\Cliff\dup \3V.
\end{eqnarray}
Thus  the Grassmann algebra 
$\3V'$ is a spinor space 
of the spin group $\spin(\dup \3V)$
of the Clifford algebra $\cliff \dup \3V$.
$\Box$

Then spin operators of one rank port spins                                                                                                                                                                                                                                                                                                                                                                                                                                                                                                                                                                                                                                                                                                                                                                                                                                                                                                                                                                                                                                                                                                                                                                                                                                                                                                                                                                                                                                                                                                                                                                                                                                                                                                                                                                                                                                                                                                                                                                                                                                                                                                                                                                                                                                                                                                                                                                                                                                                                                                                                                                                                                                                                                                                                                                                                                                                                                                                                                                  of the preceding rank, or pre-spins, 
and so on to the end, the empty set.
Addresses of $\3S$ computations  are spins.
The spins of $\3S$ are more thoroughly quantum than qubits, 
which usually have classical addresses.

I assume the converse of this proposition:

{\hypothesis [Spin] \hfil\\
\rule{6pt}{0pt} All spins in Nature are q sets.}
 \vskip8pt
I include the internal ``spins" of the Standard Model, isospin and color.
Spins are then composed of odd pre-spins, and
spin operators port pre-spins.

Following Cantor, I assume that $\iota$ can associate vectors of any grade,
although  Feynman-diagram verticess of the Standard Model have small grade,  because there is  no known bound
to the number of events in a history or a propagator, other than the
total number of events.
The Cantor assumption leads to no infinities because we use only several ranks
of $\3S$.

\section{The spinor form $\beta$\label{S:BETA}}

Let $\3P=\W \3V$ be a Grassmann algebra. 
$\3P$ is a spinor space of the group $\SO[\dup\3V]$.
To define a bilinear spinor form $\beta$ on $\3P$, 
let 
$v _1, \dots, v _{d}\in \3V$ be a basis 
and
let $v _{d+1}, \dots, v _{d+d}\in\dual  \3V$ be the dual
 (= reciprocal) basis, so that as first-grade elements of $\cliff \dup \3V$,
 they obey the Clifford postulate
\BEQ
\{v _{d+m}, v _{n}\}=\delta_{mn}\/, \quad 1\le m,n\le d\/.
\EEQ
Define the Berezin volume element
\BEQ
\left(dv \right)=dv _1\dots dv _{2d}\/.
\EEQ
Then for all $q', q\in \3P$, the {\em spinor form} for $\3P$
 is the Berezin  $B^2$ form
 \BEQ
\label{E:BEREZIN}
\beta(q',q):=\int \left(dv \right)\; \frac 12 \{q', q\}\/,
\EEQ
an odd analogue  of  the Lebesgue $L^2$ norm.
Call operators that respect the norm $\beta$, {$\beta$-\em isometric}\/.
Choose signs so that for each experimenter
import vectors have negative $\beta$-norm
and export vectors positive.
Each experimenter nevertheless has  private spaces
of import and export vectors.

\section{Indefinite metric \label{S:METRIC}}
 
In the standard quantum theory bras and kets 
both have positive norms.
When imports are not coherently superposed with exports,
 however,
their relative sign is immaterial.
On the other hand, imports and exports are superposed
in the theory of Dirac spinors
and in Bogoliubov transformations.
The Hermitian metric form $i\beta$ on Dirac spinor space $4\1C$
is positive on positive-energy spinors,
which are interpreted as particle import vectors,
and negative on negative-energy spinors,
which are interpreted as anti-particle export vectors.
Dirac 
\cite{DIRAC1974}
hinted that if (say) export vectors $\psi$ for spins have positive norms 
$\|\psi\|>0$ then 
import vectors $\phi$ could have negative norms $\|\phi\|<0$,
as though Nature allows overdrafts.
Such forms define not probabilities, 
which are positive, but system fluxes, 
which change sign under total time reversal $\2T$,
the interchange of dual import and export vectors.
In each frame, the neutral port space $\3S$ reduces to the sum of a Hilbert space $\3H$ of kets (import vectors)
and a dual Hilbert space $\dual \3H$ of bras (export vectors).

In the Dirac theory of one electron, a positive definite norm is not part of the structure of the spinor space but is provided by the experimenter.
If $v^{\mu}$ is the world velocity of the experimenter and $\beta$ is the invariant neutral spinor form, then
the positive-definite form used for probabilities by this experimenter is $P=\beta \gamma_{\mu}v^{\mu}$.
Because $P$ depends on the experimenter it is not Lorentz-invariant
in the standard sense, where the experimentee is transformed relative to the experimenter.
In the spinorial field theory of a fermion,
a probability form $P_{\sigma}$ arises that
is associated with a spacelike surface $\sigma$ 
determined by the experimenter.
This is still not an element of structure of the space.
Its invariance under Poincar\'e transformations depends
on the dynamical development of the spinor field.

Since the vectors of $\3S$ are spinors, assume that 
again the positive-definite probability norm is provided by the experimenter,
and must be shown to be invariant under transformations of the system in consequence of its dynamics.

If $|\phi\cdot \psi|^2$ is the count for the transition $\psi\to \phi$
then it vanishes if $\phi$ and $\psi$ have the same polarity:
there are no transitions between two importations or two exportations.

\section{Computational features \label{S:FEATURES}}

Ideal computations can be imagined with abstract numbers but
actual computations use existing entities
like abacus beads or thoughts of number-names.
$\3S$ computation uses quanta with port vectors in $\3S$.
One model has the following
functional organs:

The {\bf address} is a q set with
first-grade 
port vectors $\psi\in \3S(5)$,
the particle rank of $\3S$ (Table \ref{T:QSETS}).
Spin, flavor, and space-time position and momentum are part of the address.

The {\bf datum} at an address with port vactor $\psi$ 
is the binary occupation number $\psi\partial_{\psi}\doteq 0,1$.

The {\bf history} of a computation relative to a user 
has {\em history vectors} $E\in \3S(6)$.
It is the q set of the operations carried out,
also called the {\em trace}
of the computation.
 
The {\em control} of an $\3S$ computation 
has
a {\em dynamics dual vector} $D\in \3S(\4R)$  that
gives a probability amplitude 
\BEQ
A=D\circ E
\EEQ
for every history vector $E\in \3S(\4R)$.
$D$ is a reformed Dirac-Feynman
history-amplitude for the quantum computer.
It is typically a tensor of enormous grade
giving the port operations in the computation,
ordinarily combined into operators
of propagation and interaction.

\section{Quantification\label{S:QQ}}

As is well known, there is no second quantization in modern quantum field theory;
only a passage from one-quantum theory to many-quantum theory, 
an instance of the general process 
named {\em quantification} by the logician William Hamilton in 1850.
The idea of ``second quantization" is another relict of 
the state-based competitor to quantum theory.
But this early mistake has left a scar that
causes no numerical error but obscures physical meaning:

Expand a one-system port vector $\psi$ and an operator $H$ in a basis $\{e_n\}\subset \3V$ 
with dual basis $\{\6e^n\}\subset \dual \3V$:
\BEQ
\psi = \psi^n e_n\/, \quad H= e_{n'}H^{n'}{}_n\6e^n
\EEQ
In ``second quantization" it is imagined 
that the amplitudes $\psi^n$ are dynamical
variables, to be replaced by operators as in quantization.
This is consistent with the early misconception that
$\psi$ is a material wave, whose
amplitudes would indeed
be dynamical variables.

The transition from one to many odd quanta, however, 
changes the algebra of the import vectors $e_n$ and their dual export vectors $\6e{}^n$,
 not of their coefficient amplitudes, in order to allow
 a plurality of quanta.
Quantification with odd statistics imbeds the one-system import vector space $\3V$ as the first grade
 in the Grassmann algebra $\W\3V=\22^{\3V}$; 
  I  omit an associator $\iota$ for perspicuity.
The import vector $e_n \in \3V$ becomes 
the import operator $\2e_n:=e_n\w \dots$ 
of left Grassmann multiplication
by $e_n$,
and a dual export vector
$e^n$ becomes the export operator $\6{\2e}{}^n$
of left Grassmann differentiation with respect to $e_n$:
\BEQ
\6{\2e}{}^n :=\frac{\partial}{\partial {e_n}}, \quad \{\6{\2e}{}^{n'},\2e_n\}=\delta^{n'}\!{}_n\/.
\EEQ
Observables operate on the Grassmann algebra $\W\3V$
and therefore belong to the Clifford algebra $\Cliff\dup\3V$.

As a result of this confusion, a superfluous duality crept in.
``Second quantization" is said to convert an import vector (ket)
into an export operator (annihilation operator).  
Actually quantification converts an import vector $e$
in  a Grassmann algebra to an import operator 
$\2e=e\w\dots$ in a Clifford algebra.

A ``second quantization" $\6{\psi}H\psi$ looks like an expectation value but is actually an operator.
A quantification $\2e H \6{\2e}$ is also an operator, but it looks like one.

Quantification is a direct process.
Quantization analyzes a macroscopic entity
into many
microscopic quantum entities. 
It is
an inverse process, inverse
to quantification-followed-by-classical-limit.
The term ``second quantization" 
belies this inverse relation too.

The (additive) {\em quantification} of any one-quantum operator 
$H=e_{n'} H^{n'}{}_ne^n$
is the many-quantum operator
\BEA\label{E:Q}
\sum H&=&\2e_{n} H^{n}{}_{n'} \2e^{n'} \cr
&=&\2e_{n} H^{n}{}_{n'} \partial_{e^{n'}}\cr
&=:&\2e H \partial_{\2e}\cr
&=:&\2e H{}\6{\2e}.
\EEA
Here boldface indicates elements of the Clifford algebra
of a GRassmann algebra.
The duality sign is on the initial (right-hand-side) $e$,
not the final one as usual.
The quantification of a projection operator $H$ is a system-number operator $\sum H$.
Iterated quantification or multiquantification connects any rank $r$ to any rank $r'>r$
and is written as
\BEA
J(r')&=&\sum^{(r')}_{(r)} J(r) \cr
&=&\2e J(r) \6{\2e}, \cr
\2e &=&\2e(r'-1)\2e(r'-2)\dots\2e(r).
\EEA
 Each quantification stage brings in its own Clifford elements 
 $\2e(r), \6{\2e}(r)$.
$J(r)$ is called the {\em seed} of its quantification $J(r')$.

$\3S$ offers two possibilities for the duality that raises and lowers indices in quantification, the duplex form and the spinor form.
Bose quantification uses the duplex form of its Hilbert space,
but fermion quantification uses the spinor form, 
which I therefore adopt for $\3S$.
Then import operators appear both in $\2e$ and $\6{\2e}$,
and if their contributions are not to cancel
the operator $J$ being quantified must have the symmetry of the form.
This is the familiar origin of the spin-statistics theorem.

\section{Binary fields\label{S:FIELDS}}
A classical field with domain $X$ and range $Y$ has state space
$Y^X$. 
I define the quantum binary field with domain port space $\3X$
and binary range 2 to have the port space $2^{\3X}$,
the Grassmann algebra over $\3X$.
For example, if $x, x',x''$ are orthogonal spinors,
the spinor $x''+x'\w x$,
as a port vector for a binary field, is a superposition
of (a port vector for) a field with the value 1
on $x''$ and 0 elsewhere,
and a field with the value 1 on $x$ and $x'$ and 0 elsewhere.

More generally, since $2^{(XY)}=(2^X)^Y$,
the Grassmann algebra $2^{\3X\3Y}$ can have
as limit a classical field with
range $2^{X}$ and domain $Y$.

These are fermionic or odd fields.
The skew-symmetry of the Grassmann product expresses the
Pauli Exclusion Principle and Fermi-Dirac statistics.
I express even fields as composites of odd fields.
The Palev statistics of Section \ref{S:PALEV} makes this feasible.

\section{Palev statistics\label{S:PALEV}}

For any semisimple Lie algebra $A$, by {\em Palev $A$-statistics} 
I will mean that defined by the port-operator commutation relations
\BEQ
[\psi_{n''},\psi_{n'}]= c_{n''n'}{}^n\psi_n
\EEQ
where $c$ is the structure tensor of $A$ \cite{PALEV1977,PALEV2002}.
In Palev's work the algebra $A$ is simple;
the generalization to semisimple is necessary for Segal stability.
The canonical commutation relations establish a duality between
import and export operators.
The Palev commutation relations establish a triality of 
two operators relative to a third, their commutator.

{\proposition  A pair of dual odd $\3V$-quanta, 
with port vector space $\dual\3V\w \3V$,
has exact Palev statistics,
with Lie algebra $\spin \dup \3V$.}\\
  \strut\\
\noindent{\bf Proof:} The import of a pair of two odd quanta
is the Grassmann product of their individual imports.
This is a
second-grade
element of $\Cliff\dup \3V$.
The second grade of  $\Cliff\dup \3V$ generates $\spin \dup \3V$.
{$\Box$}

Palev statistics has Bose statistics as a singular limit
with a self-alignment:
\BEQ
[\0{FD}]\w[\0{FD}] \cong [\0{Palev}] \oar [\0{BE}].
\EEQ
In this limit, one moment (meaning angular momentum, or infinitesimal generator) $J_{N, N+1}$ is centralized and 
identified with a
large multiple of $i$:
\BEQ
J_{N, N+1}\oar Ni
\EEQ 
This breaks the symmetry under the 
infinitesimal ``rotations" $J_{Nn}$ and $J_{N+1, n}$,
which become canonically conjugate
Bose variables in the singular limit $N\to \infty$
\cite{FINKELSTEIN2013}.

In the theory
of fermionic liquids too,
a second-grade element of a Clifford algebra is centralized, 
reducing the orthogonal group to a canonical group
in a singular limit \cite{MATTIS1965}.
In quantum liquid theory, it is
the canonical singular limit that is taken to be 
the physical theory.
In $\3S$ theories, the finite
 $\spin(N,N)$ theory  is supposed to be the physical one.
In quantum liquid theories, space is often latticed,
breaking Lorentz invariance.
In {$\3S$ models}, 
space-time, along with its material content,
is composed of $\spin(3,3)$-invariant fermionic cells,
which are Lorentz- but not Poincar\'e-invariant.

Since the fermionic algebra is a Clifford algebra,
its second-grade elements form the Lie algebra 
of the orthogonal group of the theory of spin.
A passage from spin Lie algebra to canonical Lie algebra
occurs in Yang quantization,
Tomonaga liquids,  Palev statistics,
and, in the sequel, a modular quantum space-time $\3Y$.

\section{Summary\label{S:SUMMARYI}}

Physical processes are commonly represented as networks of
relativistic quantum creations and annihilations.
This overlooks important possibilities of modular architecture;
as did Darwin.
I set up an algebraic language $\3S$
with such an architecture
and trim the Standard Model and General Relativity to fit.
The elements created and annihilated are spins in various dimensions obeying the Pauli Exclusion Principle.
An assembly of such spins is also a spin.
Such a model has advantages:
It is stable and finite,
free of the unobservable space-time continuum,
yet consistent with the observed continuous symmetries
and complementarities of Nature;
and it is based on the operational input-output logic of quantum physics,
rather than the state-based logic of classical physics.
It suggests a natural quantum unit of the gravitational field
based on quantum complementarity.
And it provides an alternative to String Theory and Supersymmetric Quantum Theory.

The processes of  $\3S$ may be regarded as quantum computations.
Their address, memory, and program structures
are
 \BIT
 \item relativistically invariant
 \item modular, hence cellular 
\item multiordinal (involving several ranks at once)
\item local (coupling only cells with nearly the same address)
\item autogenic
\EIT
The data are quantum
binary and the addresses
quantum hyperbinary (Table I).
A  first-grade $\psi\in \3S$ 
statistically represents
the input of a cell address.
The quantum content of the cell 
is its occupation q number 
$\psi\,^{\partial}\!{\psi} \doteq 0$ or 1.
Like all partial derivatives, this Grassmann derivative 
$\partial_{\psi}$
is defined relative to an implied frame.

The algebras of
\BIT
\item Canonical commutation relations
\item Bose-Einstein statistics
\item Hilbert space 
\item Standard Model (hypercharge is its radical)
\EIT
have been physically regularized.

In {$\3S$ models}, each divergence problem of the canonical theory 
becomes a finite experimental prediction of the theory.

Anomalies are now symmetries lacking in the cell but arising
in a singular limit of many cells.
The centrality of $i$ is such an anomaly.

\end{document}